\documentclass[journal=jacsat,manuscript=article]{achemso}

\usepackage{chemformula} 
\usepackage[T1]{fontenc} 
\usepackage{soul}

\author{Yahel Soffer}
\affiliation{Department of Physics of Complex Systems, The Weizmann Institute of Science, Rehovot 76100, Israel}
\alsoaffiliation[Second University]
{These authors contributed equally to this work}
\author{Dekel Raanan}
\affiliation{Department of Physics of Complex Systems, The Weizmann Institute of Science, Rehovot 76100, Israel}
\alsoaffiliation[Second University]
{These authors contributed equally to this work}
\author{Dan Oron}
\affiliation{Department of Physics of Complex Systems, The Weizmann Institute of Science, Rehovot 76100, Israel}
\email{dan.oron@weizmann.ac.il}

\title[An \textsf{achemso} demo]
  {Low Frequency Collinear Pre-Resonant Impulsive Stimulated Raman Spectroscopy}

\abbreviations{IR,NMR,UV}

\begin{document}







\begin{abstract}
 In this work we extend low frequency impulsive stimulated Raman microspectroscopy to the pre-electronic resonance regime. We use a broadband two color collinear pump probe scheme which can be readily extended to imaging. We discuss the difficulties unique to this type of measurements in the form of competing resonant two-photon processes and the means to overcome them. We successfully reduce the noise which arises due to those competing processes by eliminating the detected spectral components which do not contribute to the vibrational signature of the sample though introduce most of the noise. Finally, we demonstrate low-frequency spectroscopy of crystalline samples under near-resonant pumping showing both enhancement and spectral modification due to coupling with the electronic degree of freedom.
\end{abstract}

\section{Introduction}
Raman spectroscopy has been, since its discovery nearly a century ago, a useful tool in the study of vibrations in matter. In contrast to fluorescence, Raman scattering provides a unique fingerprint for a particular molecular structure. Yet, the former is nowadays much more commonly used in microscopy owing to the orders-of-magnitude larger cross-section of the fluorescence of a typical emitter (7-9 orders of magnitude \cite{Wei2018}) as compared with inelastic scattering, resulting in a much lower integration time per pixel \cite{Lichtman2005,Sanderson2014}. As an attempt to enhance the Raman scattering cross-section, schemes which involved an electronically-resonant excitation were developed in the 50's and 60's. These  demonstrated enhancement of the vibrational scattering cross-section \cite{Maier1960,Strommen1977} by exploiting the electronic-vibrational coupling which promotes the nuclear motion. These methods, however, are often hampered by the presence of a large fluorescent background which overwhelms the vibrational signal \cite{Alajtal2010} and often result in sample damage. In parallel to these techniques, and mostly after the invention of ultrafast pulsed lasers, stimulated Raman scattering techniques were developed, which typically use a narrow-band two-color scheme for the selective excitation and detection of the vibrational signals. These methods rely on matching the beat frequency of the input fields to the vibrational frequency of the sample, and can generally be divided into two schemes: Stimulated Raman scattering (SRS) techniques\cite{Bloembergen1967,Boyd2003}, where the modulation transfer from a high frequency input field to a lower frequency input field is detected, and Coherent anti-stokes Raman scattering (CARS) \cite{Minck1963,Begley1974}, where the generation of new frequency components is measured. Both techniques can be interpreted as characterizing the phase modulation of the input fields due to a coherent temporal oscillation of the refractive index. Such techniques and their derivatives were very much improved in the last two decades, leading to their exploitation in microscopic imaging \cite{Zumbusch1999}, and providing vibrational-selective in-vivo images of biological samples in the C-H O-H bonding regime \cite{Freudiger2008} (around $3000cm^{-1}$) , in the fingerprint regime ($200-1500cm^{-1}$) \cite{Suhalim2012}, and recently also in the low-frequency regime  ($<200cm^{-1}$) \cite{Ren2019}. More recently, Wei et al. \cite{Wei2017} have demonstrated that a combination of a narrowband two-color scheme with a near electronic resonant pump, can benefit Raman microscopy in the fingerprint spectral region from both an outstanding molecular specificity and near-resonance enhancement due to the proximity to an electronic resonance. Notably, implementation of coherent Raman scattering for microscopy typically requires a collinear configuration, in contrast with the folded geometries often utilized in molecular spectroscopy experiments.
	
The low-frequency vibrational spectrum ($<200cm^{-1}$) is rich in information about the structural conformation of the sample of interest, including collective motion of multiple atomic and molecular groups within the molecule or unit cell, as well as relative motion between neighboring structures\cite{Parrott2015,Fischer2002,Nims2019}. While providing complementary information to that given by the high frequency regime, research in the low frequency regime has been sparse for many years. The reason for this was originally the limited spectral sharpness of edge or notch filters that are required to chromatically separate the excitation field and the scattering signal, which are very close in frequency if tuned to match a low-frequency vibrational mode. With the advent of filter technology, it became clear that there are strong background contributions due to incoherent motion without a restoring force (e.g molecular rotation) which make this spectral region difficult to observe. Recently, however, several time-domain, pump-probe techniques demonstrated the ability to detect these low-frequency vibrations by either measuring the broadband-equivalent of SRS in the form of a spectral shift \cite{Domingue2014,Raanan2018a,Weiner1991,Ruhman1987}, a CARS-like signal in the form of new frequencies that are generated from a spectrally clipped pulse\cite{Ogilvie2006}, OHD-OKE \cite{Smith2002} and Raman-induced Kerr \cite{Raanan2018}, all using oscillator-power levels which can be used for microscopy. In such time domain techniques there are no distinct pump and Stokes spectral components like in conventional two-color stimulated Raman measurements. Rather, a short, spectrally broadband, pump pulse, enables the coherent activation of all the Raman modes with a period longer than the pulse duration.
The transient oscillations of the refractive index which arise due to the vibrating molecules, then affect a delayed probe pulse, which undergoes phase modulation. If a narrowband probe is selected from the broadband pulse \cite{Ren2019}, this gives rise to spectrally narrow sidebands shifted by the vibrational frequency. If the probe pulse is broad enough and transform limited, these spectral sidebands spectrally overlap with the probe pulse and thus give rise to an overall spectral shift. Consequently the center of mass of the spectrum of the probe depends on the pump-probe delay. In off resonance impulsive stimulated Raman scattering (ISRS) this is usually measured by a spectral edge filter which transmits about 50\% of the probe spectrum and converts the spectral shift to power modulation \cite{Weiner1991,Domingue2014} or by measuring the change in the center of mass of the spectrum using a spectrometer equipped with a position sensitive detector\cite{Raanan2018a}. Alternatively, the Raman induced change in the refractive index gives rise to a lensing effect which can be detected in a manner similar to a closed-aperture z-scan \cite{Raanan2018}.

\begin{figure}[H] 
\centering\includegraphics[width=13cm]{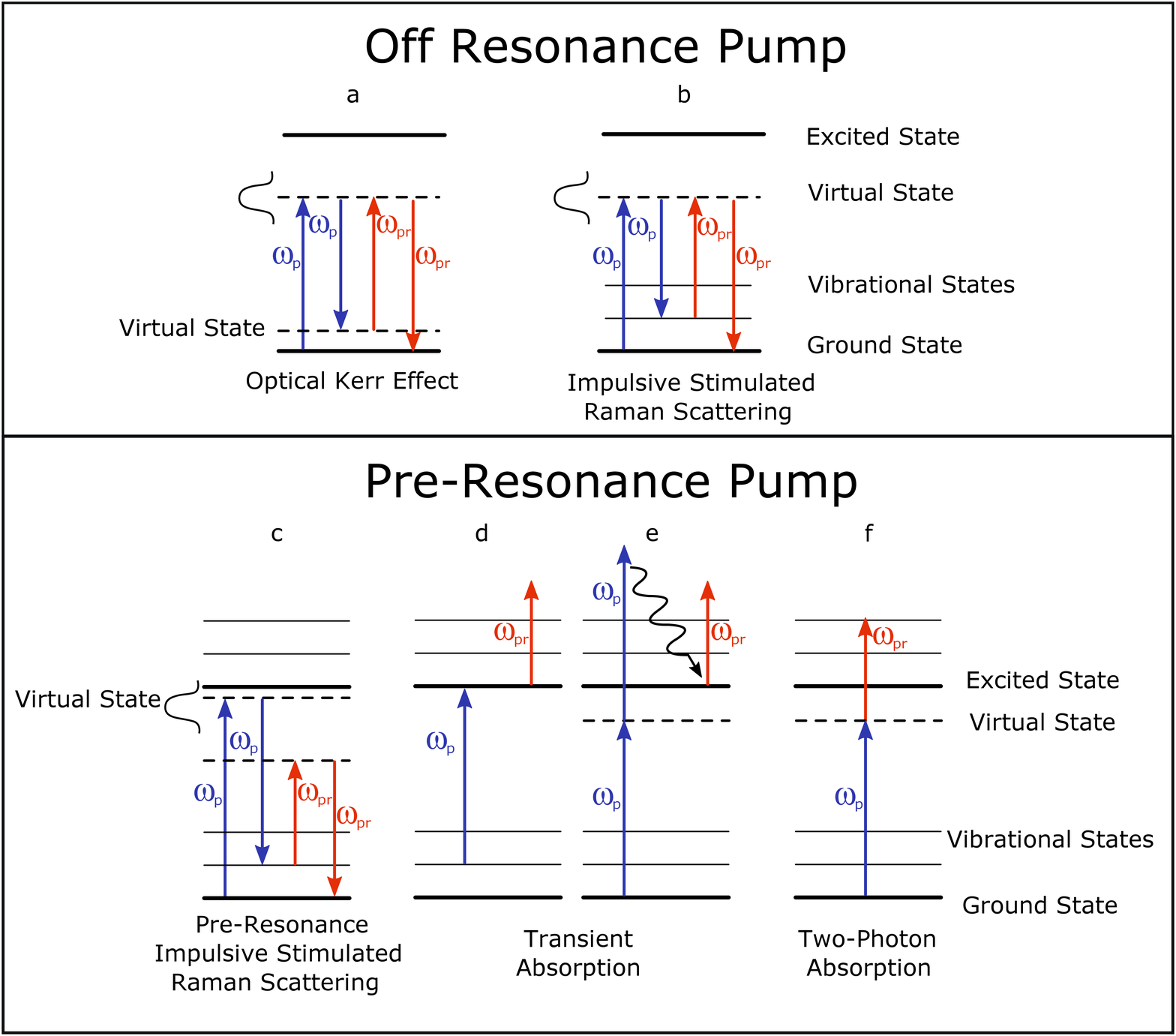}
\caption{Processes that contribute to a cross-modulation signal in two Raman configurations using an off-resonant probe. The upper and lower parts are processes which involve off- and pre-resonance pump excitation respectively. The blue arrows denote pump transitions, and red arrows denote probe transitions. (a) Optical Kerr effect (b) Impulsive stimulated Raman scattering. (c) Pre-resonant impulsive stimulated Raman scattering. (d) and (e) transient absorption of the probe after linear or two photon absorption of the pump. (f) Instantaneous pump-induced probe absorption.} 
\label{Transitions}
\end{figure}

In the resonant variant of ISRS the driving field is an electronically resonant broadband pulse. In analogy with spontaneous resonant Raman scattering where fluorescence is a major source of background, in RISRS the proximity to resonance gives rise to competing nonlinear effects which can mask the Raman signal. While working off resonance the main noise mechanism is shot noise, in RISRS these competing processes, which can be a much more significant noise source, must be dealt with.
This difference is illustrated in Figure \ref{Transitions}. When the pump is off-resonant, as depicted in figures \ref{Transitions}a and \ref{Transitions}b, the only additional mechanism coupling the pump and the probe besides the Raman process is the nonresonant Kerr effect (Figure \ref{Transitions}a) . Since this response is instantaneous it has a minor influence on a delayed probe and on the vibrational signal. In contrast, as the pump frequency approaches an electronic resonance, as depicted in figures 1c-f, several additional processes take place in addition to RISRS. In particular they include transient absorption of the probe due to  either linear (Figure \ref{Transitions}d) or two-photon (Figure \ref{Transitions}e) absorption of the probe, as well as an instantaneous pump-induced absorption of the probe via a virtual state (Figure \ref{Transitions}f). Hence the modulation-transfer signal which is usually used to isolate the vibrational signal from the noisy environment contains noisy components which are not due to the excited vibration (Figure \ref{Transitions}c).

Most previous work on RISRS was performed using a non-collinear configuration which makes it unsuitable for vibrational microscopy and microspectroscopy \cite{Chesnoy1988}. Moreover, optical excitation on resonance often results in sample damage and therefore is typically applied on a flowing sample to avoid sample bleaching \cite{Chesnoy1988,Fragnito1989}. Here we attempt to harness the potential of pre-resonant ISRS to perform low frequency vibrational microspectroscopy of organic crystals in a collinear geometry which can readily be extended to other systems. The potential advantage we introduce here is twofold. In contrast to many of the past work \cite{Chesnoy1988,Fragnito1989,Cerullo2000}, we introduce an off-resonant probe (rather than a degenerate pump and probe) which is then only sensitive to the Raman-induced Kerr effect and to pump-induced absorption (that is, probe absorption is not directly modulated by the molecular vibration). Additionally, we identify a way to increase the signal to noise in pre-resonant pump conditions, as we observe the appearance of new background and noise sources in such conditions. Our solution, based upon a judicious choice of the fraction of probe frequencies to be measured, enables us to clearly extract low-frequency signatures even in the presence of significant absorption in the sample in a geometry which can be simply adjusted to microscopy. We show that such a scheme can provide information about collective motion of molecules under electronic-resonance conditions and reveal vibronic coupling. 

\section{Results and discussion}

\begin{figure}[H] 
\centering
\centering\includegraphics[width=12cm]{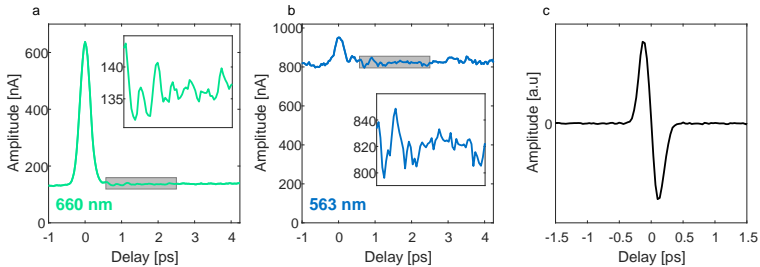}
\caption{Modulation transfer signals using (a) 660nm pre-resonance pump and (b) 563nm on resonance pump wavelength, normalized by the AOM modulation ratio. The BP filter is aligned to clip 50\% of the spectrum in both (a) and (b). The darkened regimes in (a) and (b) are plotted in the insets. (c) Optical Kerr effect in glass.}
\label{Tetracene660_563_exc_abs}
\end{figure}

As a demonstration of our method we measured organic single crystals of tetracene and orthorhombic rubrene, both containing a condensed benzene ring structure, using near and on resonance pumping. In figures \ref{Tetracene660_563_exc_abs}a and b we show typical vibrational time traces of tetracene using the same conditions under two different excitation wavelengths: 660nm, where the absorbance of the tetracene crystal is weak (figure \ref{Tetracene660_563_exc_abs}a), and 563nm, which already strongly overlaps with the onset of the crystal absorption (figure \ref{Tetracene660_563_exc_abs}b). Two features of the spectrum clearly stand out: A strong absorptive signature at zero delay, and a large, seemingly flat background (note that the background is practically equal before and after zero delay), indicating long-lived absorption with a lifetime much greater than the 12ns repetition rate of our laser. To better understand the origin of the peak at zero delay, we can first compare it with the non-resonant Kerr response of glass, shown in Figure \ref{Tetracene660_563_exc_abs}c. Since the pulses temporal intensity profile can be reasonably approximated by a Gaussian, the Kerr effect results in a signal with the shape of a Gaussian derivative. It stands out that the zero time delay response of tetracene in both excitation wavelengths has the shape of an isolated peak and does not resemble that of glass. It must therefore arise from a resonant effect. From Allan deviation measurements (see supporting information) we infer that the enhanced noise associated with pumping tetracene at 563nm is due to internal processes in the sample, rather than laser-noise. The most reasonable association is pump-induced absorption as depicted in figure \ref{Transitions}f: When the pump and probe overlap in time, simultaneous absorption of one pump photon and one probe photon is enabled. This effect clearly overwhelms the much weaker Kerr modulation. The long-lived absorption component is likely due the accumulation of excited carriers (more likely trapped carriers) in the tetracene matrix, associated with the absorption processes denoted in figures \ref{Transitions}d and \ref{Transitions}e. This flat background limits the measurement in two ways. It reduces the dynamic range of the detector, and introduces a strong noise, probably due to fluctuations in excited carrier density, which can completely overwhelm the Raman signal. This calls for a means for significant reduction of this noise component. 

\begin{figure}[H] 
\centering\includegraphics[width=11cm]{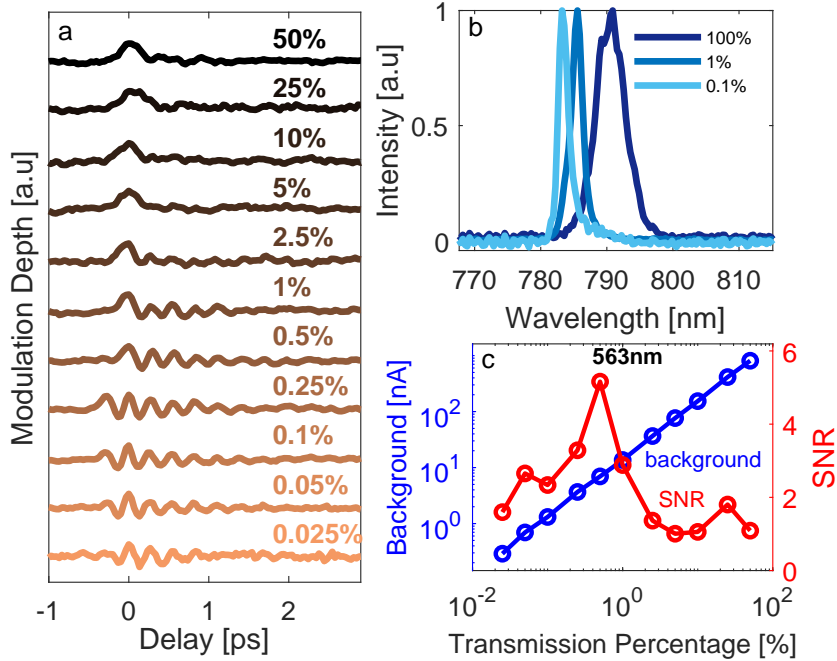}
\caption{(a) ISRS measurement after background subtraction and normalization at 563nm excitation wavelength at different probe spectral windows, manifested as different transmission percentage. The oscillation depth is at the Nano-Ampere scale. (b) Normalized probe spectrum at the detector at different angles of incident at the BP filter (different fraction of probe transmittance). (c) Lock in output background (blue) and SNR (red) as a function of the transmission percentage at different spectral windows with pump wavelength of 563nm. The SNR of the measurement represents the ratio between the vibrational peak of tetracene at $120cm^{-1}$ and the standard deviation of the measurement at low frequencies excluding the $50cm^{-1}$ band. Notably, in order to remove slowly-varying drifts, a polynomial fit of rank 3 was subtracted from the time-domain trace for the calculation of the SNR.}
\label{Tetracene_angles}
\end{figure}

We successfully handle this problem by blocking a major part in the probe power spectrum. In figure \ref{Tetracene_angles}a we show the time trace of the vibrational signal as a function of the fraction of the energy of the probe transmission through the band-pass (BP) filter (at a pump wavelength of 563nm, close to the onset of the absorption band). The top presents the data shown in figure \ref{Tetracene660_563_exc_abs}b and does not show a clear vibrational signature. As we rotate the BP filter (see experimental setup) the transmitted spectrum is blue shifted from the probe central frequency, as seen in figure \ref{Tetracene_angles}b and less probe energy is transmitted. This results in clear observation of a delayed ringing associated with stimulated vibrations, at a much higher signal to noise ratio, as seen in figure \ref{Tetracene_angles}a,c. We note that for the calculation of SNR presented in figure \ref{Tetracene_angles} c we subtracted a polynomial fit of rank 3 from the time trace between 570 fs and 4.5 ps in order to eliminate the effect of slowly-varying drifts on the analysis. Notably, a maximal value of the SNR is achieved at a transmittance of around 0.5\%. This corresponds to a tradeoff between noise from transient absorption, dominating at high transmission, and detector noise, dominating at low transmission.

\begin{figure}[H] 
\centering\includegraphics[width=11cm]{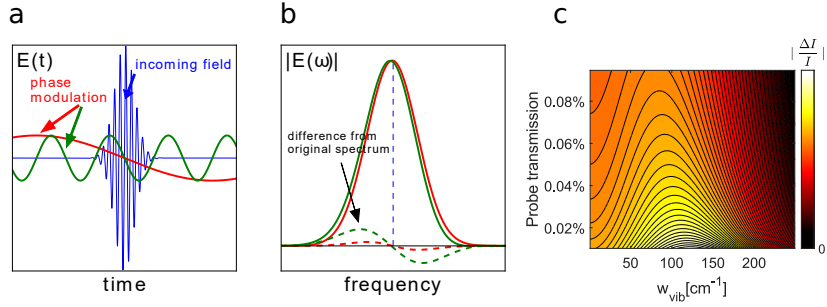}
\caption{(a) An ultra-short pulse is phase modulated at a high frequency (green) and low frequency (red) due to an oscillating refractive index. (b) The phase modulation exerted by the oscillating in time refractive index affects the spectral components of the incoming field. The deviation of the spectrum (dotted) is maximized  at approximately one vibrational frequency away from the carrier frequency, and occurs farther away from the carrier frequency for the high freqency modulation (green) as compared to the low frequency modulation (red). The blue dotted line represents the central frequency of the original spectrum. (c) Simulation of the relative modulation transfer as a function of the vibrational frequency and the transmission through the clipping spectral filter.}
\label{PhaseModulation} 
\end{figure}

In order to explain this observation, we simulated the effect of the clipping of the spectrum on the ratio between the cross modulation amplitude and the bias. As illustrated in figure \ref{PhaseModulation}a and figure \ref{PhaseModulation}b, the strongest contribution to the modification of the probe spectrum by the vibrationally induced phase modulation occurs at a frequency separated about one vibrational frequency away from the carrier frequency, while the modification of the spectrum around the central frequency is minimal. In contrast, pump-induced absorption of the probe affects the intensity of the whole spectrum. Consequently our spectral shift signal is observed on top of a large bias which comes along with noise levels proportional to this bias. Thus frequency components close to the central frequency of the probe pulse introduce higher background. In order to increase the signal to background (and signal to noise) we crop the spectrum and detect only a chosen window at the edge of the probe's spectrum. This way we eliminate the spectrally non-contributing components at the central part of the spectrum which limit the measurement dynamic range and introduce most of the noise.

We retrieve the vibrational spectrum by performing Fourier transformation of the measured output of the lock-in amplifier on a time window extending from about 500-600fs following zero delay to 4-5ps (this translates to about 8{$cm^{-1}$} spectral resolution), although we note that there are alternative methods which are possibly preferential in terms of SNR \cite{Sinjab2020}. We mention that while the noise sources discussed in this section are not related to shot-noise, the same arguments are also valid for the latter as it also depends on the input power, and doesn't depend on the wavelength. In figure \ref{PhaseModulation}c we present a numerical simulation, that takes into account the induced dipole effect and uses the Lorentz model to calculate the spectrum of the probe pulse as a function of the frequency of vibration. We assume the same transition strength for all vibrational frequencies and approximate a Gaussian temporal envelope for the pump and probe fields. We then plot the relative power modulation as a function of both the frequency of vibration and the fraction of energy transmitted through the detected spectral window. The color scale denotes the ratio between the magnitude of the modulation transfer power and that of the background power that reaches the detector. While the simulation shows that there is no cutoff wavelength which results in the highest signal-to-background ratio, such an optimal cut-off frequency must exist in the presence of experimental noise, be it detector noise or shot noise.

\begin{figure}[H] 
\centering\includegraphics[width=11cm]{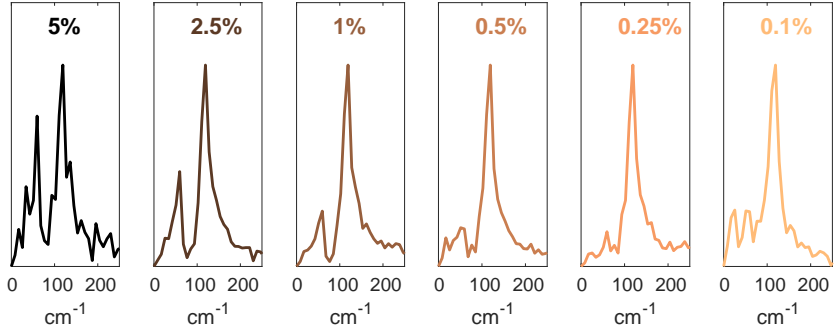}
\caption{Extracted Raman spectrum at different clipping windows of the probe spectrum, pumping at 660nm. A notable trade-off exists between high sensitivity and the ability to detect low frequency vibrational modes as we crop the spectrum at a higher cut-off frequency. Further cropping below 2.5\% introduces noise originating from detector noise and shot noise which dominate the noise in the measurement at low power.}
\label{SpectraVsTransmittance}
\end{figure}

While we demonstrate improvement of SNR as we clip the spectrum of the probe, a notable trade-off exists in the form of a reduction of signal in the lower frequency range. As predicted qualitatively by the simulation in figure $\ref{PhaseModulation}$c and presented in figure  \ref{SpectraVsTransmittance}, a gradual reduction of the lower frequency end occurs as we push the cut-off frequency away from the carrier frequency. Furthermore, as can be seen in figure \ref{Tetracene_angles}c, the optimal experimental value in terms of SNR at 563nm pumping was 0.5\%. Notably, this depends on the pump wavelength, and for a pump wavelength of 660nm see figure \ref{SpectraVsTransmittance}, we found that the optimum clipping is around 2.5\% transmission which maintains the low frequency band at $50 cm^{-1}$ visible while significantly improving the SNR. 

\begin{figure}[H] 
\centering\includegraphics[width=15cm]{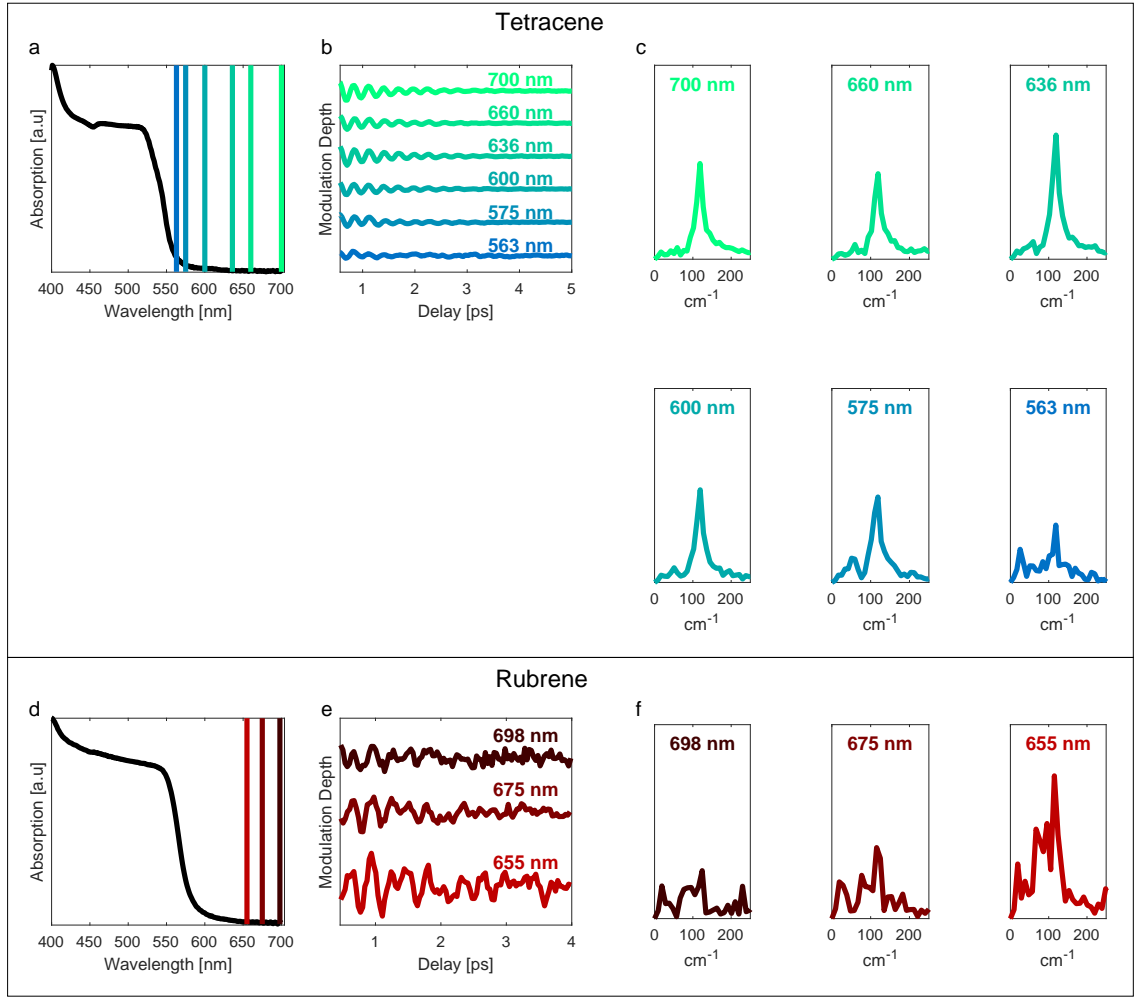}
\caption{The effect of approaching resonant conditions on tetracene and rubrene ISRS spectra. (a) Absorption spectrum of the tetracene sample. Vertical lines are plotted at the different excitation wavelengths. (b) Tetracene ISRS signal at different excitation wavelengths at transmission of 0.25\% of the probe power through the BP filter, after background subtraction and normalization by the AOM modulation ratio. (c) Fourier transform of the oscillations at the different pump wavelengths after subtraction of a third order polynom to remove slowly varying drifts. (d) Absorption spectrum of rubrene. Vertical lines are plotted at the different excitation wavelengths.(e) Rubrene ISRS signal at different excitation wavelengths at 0.25\% probe transmission, after background and polyfit subtraction normalized by the AOM modulation ratio.(f) The vibrational spectrum acquired with different excitation wavelengths at transmission of 0.25\% of the probe power through the BP filter. The scale of all subplots in (c) and (f) is identical.}
\label{TetraceneDiffWavelength}
\end{figure}

In figure \ref{TetraceneDiffWavelength} we present the ISRS signal of tertracene  and rubrene at 6 and 3 excitation wavelengths, respectively, close to and in resonance. Saturation effects were not observed for all pump wavelengths, mostly owing to the pre-resonance scheme. As discussed above, measurements for each material were done with similar pump and probe powers and pulse duration, and were normalized by the AOM modulation depth. In the tetracene spectra we observe a gradual increase in magnitude of the $50cm^{-1}$ band as we excite at shorter wavelengths towards the resonance, as shown in figure \ref{TetraceneDiffWavelength}c. We note that even upon 563nm excitation, at 10\% of the absorption at 500nm, we can still identify ringing in the time trace, and a corresponding peak in the Raman spectrum just above the noise level. At this proximity to resonance the SNR is very much reduced as compared to a pre-resonance excitation, due to both additional noise and weaker vibrational signature. The variation of the Raman spectrum in rubrene sample is even more conspicuous as the Raman spectrum in figure \ref{TetraceneDiffWavelength} f shows a significant gradual increase towards pre-resonant excitation (also shown in the time domain in figure \ref{TetraceneDiffWavelength} e). We relate these to the different coupling between the electronic excitation and the different vibrational modes, resulting in different enhancement factors. The spontaneous low frequency Raman spectra of tetracene and rubrene can be found in \cite{WeinbergWolf2007,Socci2017,Venuti2004}. All measurements presented are taken using a lock-in integration time of 300ms per point with a single sweep, and a step size of 33fs.

In summary, we have demonstrated a collinear ultrafast pump-probe technique to measure low frequency vibrational spectrum when the pump wavelength approaches an electronic resonance and the probe is far detuned. It is based on the unique properties of phase modulation that is experienced by a probe pulse as it propagates through a vibrating in time refractive index. We demonstrated ISRS of crystalline tetracene and rubrene in the proximity of an electronic resonant level and successfully overcame the issue of increased noise due to other competing nonlinear processes occuring upon resonant and near-resonant conditions. We identified those as pump induced absorption and transient excitation. We surpassed this hurdle via detection of the frequency modulation of the probe pulse through a cropped window of the probe spectrum, achieving a significant reduction of the background and noise, with a corresponding increase of SNR. Notably, a tradeoff exists between high signal-to-background ratio and the ability to observe low vibrational frequencies. We witnessed changes of either the vibrational spectrum or the signal magnitude as we approach the electronic resonance. We believe this method will allow a broad use of ISRS in the vicinity of an electronic resonant level in chemical characterization, molecular identification and label-free imaging.

\section{Experimental}

\begin{figure}[H] 
\centering\includegraphics[width=13cm]{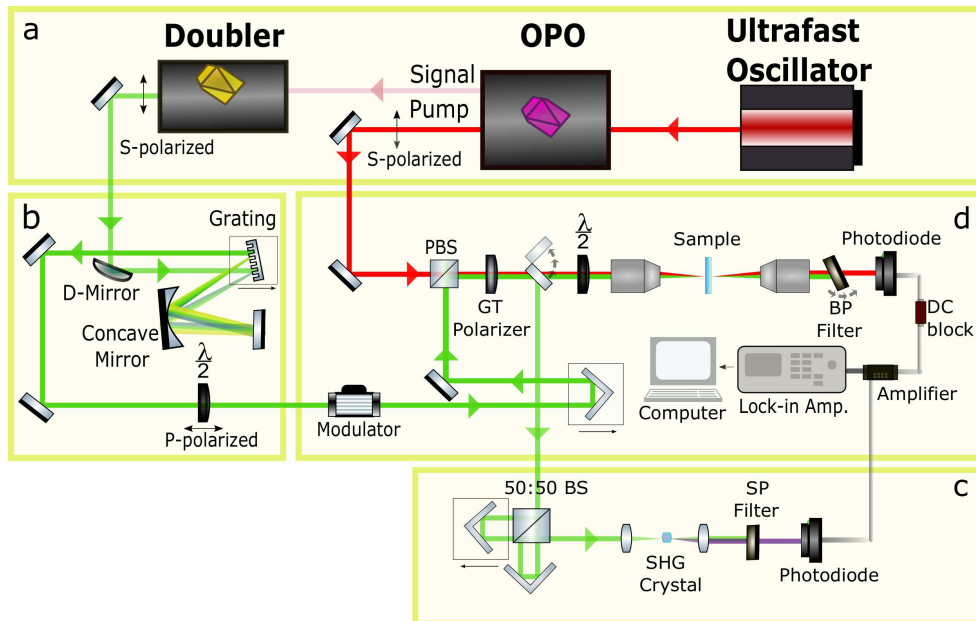}
\caption{Layout of the experimental setup: (a) Laser source. (b) Pulse compressor. (c) Interferometric autocorrelator. (d) Combining the pump and probe and detecting the probe.}
\label{ExperimentalSetup}
\end{figure}

For the time resolved resonance and pre-resonance ISRS spectroscopy we tune the pump wavelength, while keeping the probe wavelength fixed and far detuned, thus removing unwanted direct resonant effects from the detection apparatus. The experimental setup is shown in figure \ref{ExperimentalSetup}. We use an ultrafast Ti:Sapphire mode-locked oscillator (Chameleon Ultra II, Coherent) producing 120fs pulses at 80MHz with maximum average power of 3.5W. The output of the oscillator is directed into an optical parametric oscillator with a typical output power of 1W (Chameleon Compact OPO VIS Version 1.2, Coherent). The OPO residual pump output, centered at 790nm, functions as the probe beam in the experiment. The tunable OPO signal output is sent into a frequency doubling unit (Coherent), to produce a tunable second harmonic output (500-700nm , 100mW typical output power) that functions as the pump beam in the pump-probe experiment. Following each pump color change we direct the pump beam into a home-built interferometric autocorrelator to measure the duration of the pump pulse. We minimize the chirp while observing the autocorrelation trace  using a home-built pulse compressor composed of a grating (Richardson, blazed at 610nm, 1200 grooves/mm) a curved mirror with 20cm focal length and a planar mirror. We found that the probe pulse doesn't require additional compression. We then modulate the pump using an acousto-optic modulator at 86kHz. The delay between the pump and the probe is scanned by a motorized delay stage placed on the pump arm. Prior to their recombination, the pump and probe polarizations are set to be orthogonal with respect to each other and a polarizing beam-splitter is used to merge them collinearly. We then use a Glan-Taylor polarizer to align the two colors to the same polarization and rotate it using an achromatic half wave plate to overlap with the most effective component in the third order non-linear tensor of the sample. We demonstrate our technique on both tetracene and rubrene single crystals. Since working upon electronic resonance excitation in crystalline samples the power of the beams had to be carefully selected to avoid photodamage. We chose the pump and probe powers in both tetracene and rubrene to be slightly below the power which results in damage to the sample at the shortest pump wavelength that was used, and kept the powers constant while changing the pump wavelengths. The power at the tetracene sample was set to 1.1mW and 10mW for the pump and probe beams respectively. A 5mW pump and 10.8mW probe were used for rubrene. The two beams are focused on the sample using a 20X 0.33NA achromatic objective and collected in the forward direction by a 3cm focal length achromatic lens. The pump is then chromatically filtered out. A band pass (BP) filter (20nm span around 800nm) which is used as an edge filter, is positioned on a rotation stage and partially blocks the probe spectrum. The transmission window through the BP filter can be adjusted by rotation of the filter, and is quantified below by the fraction of power transmitted. The beam is then directed to a large aperture photodiode (DET100A2 Thorlabs). We observe the fine modulation transfer signal on top of a relatively strong background signal. To extract the fine cross-modulation signals we filter out the large bias using an electronic DC-block filter (EF599 Thorlabs), amplify the remaining modulated component by a low noise trans-impedance amplifier (DHPCA-100, femto), and finally measure using a lock-in amplifier (SR830, Stanford Research Systems) with integration time of 300ms. The Fourier transform of the output signal of the lock-in amplifier as a function of the pump-probe delay is the ISRS spectrum.

\section*{Funding}
This work was supported by a research grant from the Benoziyo Endowment Fund for the Advancement of Sciences ; by the Crown Center of Photonics ; by the Weizmann-CNRS Collaboration Program and by the ImagiNano European Associated Laboratory; D.O. is the incumbent of the Harry Weinrebe Professorial Chair of laser physics.

\begin{acknowledgement}
The authors thank Omer Yaffe and Maor Asher for sample preparation and their input regarding tetracene's structure, and Herv\'{e} Rigneault for helpful discussions.



\end{acknowledgement}






\bibliography{ms}
\end{document}








\section{Allan deviation measurement}

In this supporting information we analyze the timescale of noise in our setup by measuring its Allan deviation. The Allan deviation is measured both for Tetracene and glass as a reference, using both 563nm and 660nm excitation. The pump probe delay for the Allan deviation of tetracene is tuned to avoid the instantaneous response, yet still within the molecular ringing coherence time. The relative pump probe delay for the glass sample is tuned to maximize the instantaneous Kerr effect. The results, presented in figure \ref{AllanDeviation}, show that the noise with glass sample is independent of the excitation wavelength, which suggests that the enhanced noise observed for near-resonant (563nm) excitation in tetracene rises due to internal processes rather than laser noise. Results are shown both for 0.25\% transmission and 50\% transmission through the band pass filter. Each measurement was taken over 6 hours, with separation of 750ms between consecutive data points. The integration time of the lock-in amplifier is 300ms. The power levels used here for the pump and probe are the same as used throughout the measurements presented in the main text for tetracene. We mention here that the measurement duration for the data presented in figures 5 and 6 in the main text is about 1.5 minutes, suggesting that the SNR could still be improved by increasing the integration time before slow drifts govern the noise.

\renewcommand{\thefigure}{S\arabic{figure}}
\begin{figure}[H] 
\centering\includegraphics[width=11cm]{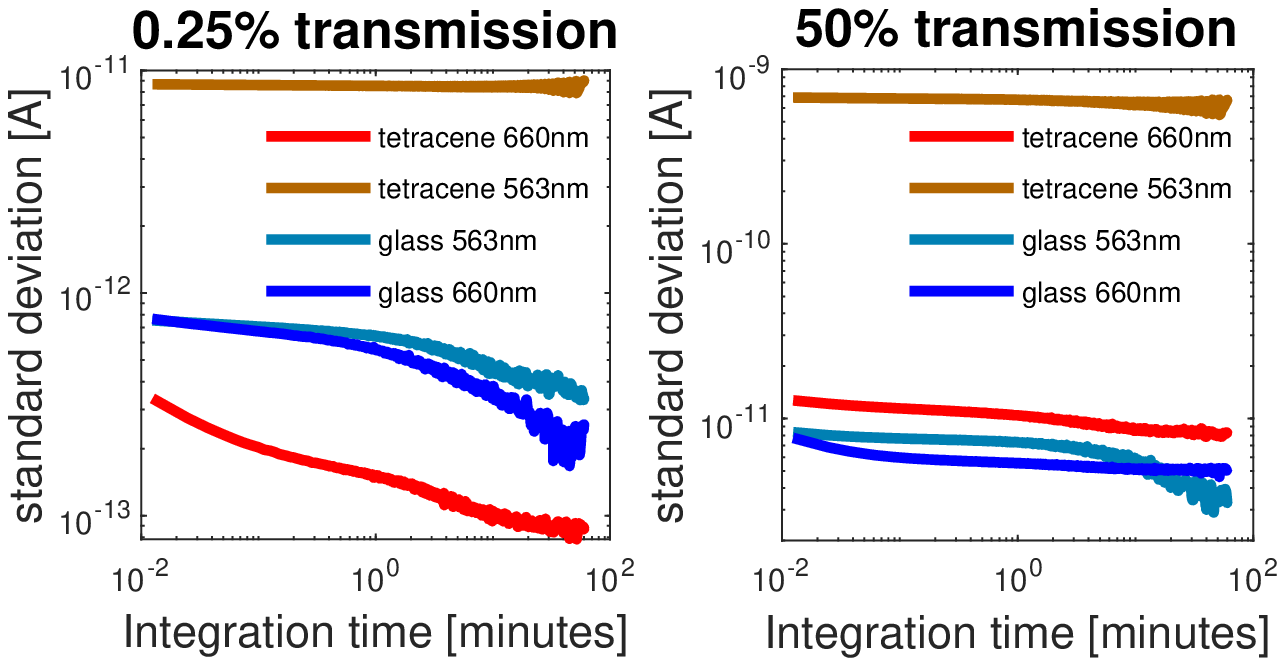}
\caption{Allan deviation measurements for both tetracene and glass, for 0.25\% (left) and 50\% (right) transmission through the BP filter. Data is presented for off-resonant pumping at 660nm and near-resonant pumping at 563nm.}
\label{AllanDeviation}
\end{figure}